\documentclass[12pt]{article}
\usepackage[dvips]{graphicx} 
\usepackage{amsmath}
\usepackage{tabularx} 
\usepackage{setspace}   
\numberwithin{equation}{section}
   
\newcommand{\ba}{\begin{align}} 
\newcommand{\ea}{\end{align}}

\usepackage{amsfonts} 
\usepackage{amssymb}

\newtheorem{thm}{Theorem}
\newtheorem{cor}[thm]{Corollary}
\newtheorem{lemma}[thm]{Lemma}

\newtheorem{prop}[thm]{Proposition}
\newtheorem{defn}[thm]{Definition}

\def\qed{{\bf QED}}

\def\tr{\hbox{Tr}}

\def\be{\begin{eqnarray}}
\def\ee{\end{eqnarray}}
\def\bee{\begin{eqnarray*}}
\def\eee{\end{eqnarray*}}
\def\bmx{\begin{pmatrix}}
\def\emx{\end{pmatrix}}

\def\ts{\textstyle}
\def\bra{\langle}
\def\ket{\rangle}
\def\kb{ \ket \bra }

\def\rt2{\ts \frac{1}{\sqrt{2}} }

\def\ot{\otimes}

\def\tu{{\rm Tube}}

\title{Entanglement of random subspaces via the Hastings bound}

\author{Motohisa Fukuda\\
{\small Department of Mathematics}\\
{\small University of California, Davis CA 95616}
\and Christopher King\\
{\small      Department of Mathematics}\\
{\small Northeastern University,  Boston MA 02115} }

\begin{document}

\maketitle

\begin{abstract}
Recently Hastings \cite{Hastings08} proved the existence of random unitary
channels which violate the additivity conjecture. In this paper we use Hastings' method to 
derive new bounds for the entanglement of random subspaces of bipartite systems.
As an application
we use these bounds to prove the existence of non-unital channels
which violate additivity of minimal output entropy.
\end{abstract}

\tableofcontents

\section{Introduction}
In his 2008 paper \cite{Hastings08} M. Hastings proved the existence of channels which exhibit
non-additivity of minimal output entropy. This result settled a long-standing open problem in quantum 
information theory. Hastings' paper is interesting from many points of view, not least because it
introduced some essentially new ideas into the field of random channels. To review the history a little,
in earlier work Hayden, Leung and Winter \cite{HayLeuWin06} had derived 
bounds for the entanglement of random subspaces of bipartite spaces (these bounds are recalled below in
Theorem \ref{thm:HLW}). They used concentration of
measure arguments to analyze the entropy of random states in high-dimensional spaces,
together with the ``$\epsilon$-net'' method to control the entropy of all states in a subspace.
Their analysis
led to the proofs by Hayden and Winter \cite{HaydenWinter08} of the existence of channels violating
additivity of Renyi entropy for all $p > 1$. Further progress in this direction appeared in the recent
work of Collins and Nechita \cite{CollinsNechita1,CollinsNechita2} on Renyi entropies of entangled states and subspaces.
However the $p=1$ case remained open until Hastings provided the new ingredients to complete this program.

Our goal in this paper is to apply these new methods from the paper \cite{Hastings08} to the 
analysis of random subspaces of bipartite
spaces. As an application we derive results about the entanglement of a generic high-dimensional subspace,
and show that in some regimes this provides strictly tighter bounds than the Hayden, Leung and Winter estimates.
We also use these bounds to deduce the existence of non-unital channels which violate
additivity of minimal output entropy, and in fact show that such violation is generic for high-dimensional channels.
In the process of deriving these results we formulate an abstract version
of Hastings' method, and we believe that this formulation will be useful for the study
of other generic properties of random subspaces. 

The idea of using Hastings' method to study
entanglement of random subspaces also appeared recently in the work of
Brandao and Horodecki \cite{BranHoro09}. Their work is particularly interesting because it uses a combination
of standard concentration of measure arguments together with some of the new ideas of Hastings.
There is some overlap between their paper and ours, and in particular
we re-derive their entanglement bound as a special case of our Theorem \ref{thm1}.
However we also extend their results in several ways, both by considering different dimensions for input and output
spaces, and by presenting explicit bounds for the size of the additivity violations.

\medskip
The paper is organized as follows. In the rest of this Introduction we recall the entanglement bounds 
derived by Hayden, Leung and Winter, and state the new bounds derived using Hastings' method.
We then use these bounds to prove the existence of a new class of channels with non-additive minimal output entropy.
Section 2 contains the main result of this paper, which is a general formulation of the 
Hastings bound. With an eye to possible future applications we state this as broadly as possible,
namely as a condition which guarantees convergence to zero of the probabilities of a sequence of events in the output space.
In Sections 3 and 4 we use the Hastings bound to derive the entanglement results in Section 1.
In Section 5 we prove the Hastings bound, using methods similar to those in the paper \cite{FKM09}.
Finally the Appendix contains some technical estimates needed for the derivations of the bounds.

\subsection{Entanglement of subspaces}
Consider a subspace $C$ of the bipartite system $A \ot B$.
The entanglement of $C \subset A \ot B$ is defined to be
\be
E(C) = \inf_{| \phi \ket} S\Big(\tr_B | \phi \kb \phi |\Big)
\ee
where the infimum runs over
normalized states $| \phi \ket$ in $C$,
$S(\cdot)$ is the von Neumann entropy, and $\tr_B | \phi \kb \phi |$ is the reduced
density matrix of the orthogonal projector onto $| \phi \ket$. Note that $E(C)\ge0$ with equality
if and only if $C$ contains a product state. 

In the search for counterexamples to additivity, one is interested in finding subspaces
with large entanglement. Thus the quantity $\sup E(C)$ is of interest, where the supremum
runs over all subspaces of a fixed dimension. This supremum depends only on the dimensions
of $A,B,C$. Let $d = {\rm Dim} \,A$, $n = {\rm Dim} \,B$ and $s = {\rm Dim} \,C$, then
the maximum
entanglement of a $s$-dimensional subspace in $A \ot B$ is
\be\label{def:Emax}
E_{\max}(s,d,n) = \sup \{ E(C)\,:\, C \subset A \ot B, \quad {\rm Dim} \,C = s \}
\ee
Hayden, Leung and Winter \cite{HayLeuWin06} obtained the following lower bounds
for $E_{\max}$.

\begin{thm}[Hayden, Leung and Winter]\label{thm:HLW}
Assume  that $3 \le d \le n$. Then
\be\label{HLW1}
E_{\max}(s,d,n) \ge \log d - c_1 \frac{d}{n} - c_2  \, \bigg(\frac{s+1}{d n}\bigg)^{2/5} \, \log d
\ee
where $c_1 \simeq 1.44$ and $c_2 \simeq 19.84$.
\end{thm}

The main results we present in this paper are new lower bounds for $E_{\max}(s,d,n)$.
The bounds are valid for sufficiently large dimensions $n$ and $s$, and for any dimension
$d \ge 2$. Theorem \ref{thm1} concerns the case where $s$ scales
linearly with $n$, and Theorem \ref{thm2} covers the case where $s/n \rightarrow 0$ as
$n \rightarrow \infty$.

\medskip
In order to state our first result we need to introduce the solution of the
following optimization problem: for $x,y > 0$  define
\be\label{def:Q}
h_d(x,y) = \inf_{0 < \gamma < 1} \,
\inf_{z > 1} \bigg\{\frac{z \log z + (d-z) \, \log \Big(\frac{d-z}{d-1}\Big)}{x \,(\gamma + (1-\gamma) \log (1-\gamma))} 
\,:\,  \frac{- \log z - (d-1)\,  \log \Big(\frac{d-z}{d-1}\Big)}{ - \log (1 - \gamma)} =  y  \bigg\}
\ee

\begin{thm}\label{thm1}
Let $d \ge 2$, $0 < r_1 \le r_2$, and $h >  h_d(r_1,r_2)$.
There is $n_0 < \infty$ such that for $n \ge n_0$,
and all $s$ satisfying $r_1 \le s/n \le r_2$,
\be\label{thm:1}
E_{\max}(s,d,n)  > \log d - h \, \Big(\frac{s}{n d}\Big)
\ee
\end{thm}

\medskip
The above result is a generic property, meaning that with probability approaching one as $s,n \rightarrow \infty$,
the right side of (\ref{thm:1}) is a lower bound for the entanglement $E(C)$ of a randomly
selected subspace $C$. It is possible to analyze the function $h_d$ in detail but for our purposes here
it is sufficient to note that it satisfies an upper bound which is uniform in $d$.
As mentioned before, using related methods Brandao and Horodecki \cite{BranHoro09}
proved Theorem \ref{thm1} in the case $r_1=r_2=1$. 

\medskip
Our second result concerns the case where $s/n \rightarrow 0$. Define
\be
h_0 = \inf_{0 < \gamma < 1} \frac{- \log(1-\gamma)}{\gamma + (1-\gamma) \log (1-\gamma)} \simeq 3.351
\ee

\begin{thm}\label{thm2}
Let $d \ge 2$, and $h > h_0$.
Consider sequences $\{s_k,n_k\}$ such that
\be
\frac{s_k}{n_k} \rightarrow 0, \quad
\frac{n_k \log s_k}{s_k^{3/2}} \rightarrow 0 \qquad
\mbox{as $k \rightarrow \infty$}
\ee
There is $k_0 < \infty$ such that for $k \ge k_0$,
\be\label{thm:2}
E_{\max}(s_k,d,n_k)  > \log d - h \, \Big(\frac{s_k}{n_k d}\Big)
\ee
\end{thm}

\medskip
Again we note that the lower bound in (\ref{thm:2}) is generic for random subspaces in high dimensions. 
The bounds (\ref{thm:1}), (\ref{thm:2}) and (\ref{HLW1}) can be compared for small values of the ratio $s/nd$.
It can be seen that the right side of (\ref{thm:1}), (\ref{thm:2}) behaves like $\log d - c (s/nd)$, while the right side of
(\ref{HLW1}) behaves like $\log d - c' (s/nd)^{2/5} \, \log d$ for some constants $c, c'$. Thus (\ref{thm:1}), (\ref{thm:2}) provide a sharper
bound in the regime $s << n d$.

\begin{cor}
Let $d \geq 2$, $h>h_0$ and $0< \epsilon < 1/2$.
Then, there exists $s_0$ such that
\be
E_{\max}(s,d,sd)>  \log d - h \, \Big(\frac{1}{ s^{ 1/2 - \epsilon } d}\Big)
\ee 
for all $s \geq s_0$.
\end{cor}
To see this, let $n= \lceil s^{ 3/2 - \epsilon }\rceil$ 
and then Theorem \ref{thm2} implies that
there exists $s_0$ such that
\be\label{boundcor}
E_{\max}(s,d,\lceil s^{ 3/2 - \epsilon }\rceil)  
> \log d - h \, \Big(\frac{s}{\lceil s^{ 3/2 - \epsilon }\rceil d}\Big)
> \log d - h \, \Big(\frac{1}{ s^{ 1/2 - \epsilon } d}\Big)
\ee
for $s \geq s_0$.
Without loss of generality, we can assume that
$d< s_0^{1/2 -\epsilon}$,
which implies $sd \leq \lceil s^{3/2 - \epsilon} \rceil$ for $s \geq s_0$.
As described in the next section, there is a correspondence between
subspaces of bipartite spaces and quantum channels.
Thus
a subspace which satisfies the bound in (\ref{boundcor}) corresponds to
some quantum channel where
the dimensions of input and output spaces and the number of Kraus operators
are $s, d, \lceil s^{3/2 - \epsilon} \rceil$ respectively.
However, this channel can be rewritten by using at most $sd$ Kraus operators 
\cite{zycBen04},\cite{RuskaiFI}.
Therefore this channel corresponds to 
some $s$-dimensional subspace, say $C$, of $\mathbb{C}^d \otimes \mathbb{C}^{sd}$,
for which $E(C)$ satisfies the same lower bound (\ref{boundcor}).

\subsection{Violations of additivity}
The subspace $C \subset A \ot B$ is defined by an embedding
$W: \mathbb{C}^s \rightarrow \mathbb{C}^d \ot \mathbb{C}^n$ satisfying $W^* W = I$,
where $C$ is the image of $W$. This embedding defines
two conjugate channels $\Phi_W$ and $\Phi_W^C$ via
\be
\Phi_W(\rho) = \tr_{\mathbb{C}^{d}} W \rho W^*, \quad \Phi_W^C(\rho) = \tr_{\mathbb{C}^{n}}  W \rho W^*
\ee
Letting $\overline{W}$ denote the complex conjugate of the matrix $W$,
the complex conjugate channels $\overline{\Phi}_W$ and $\overline{\Phi}_W^C$ are defined by
\be
\overline{\Phi}_W(\rho) = \tr_{\mathbb{C}^{d}}  \overline{W} \rho \overline{W}^*, \quad
\overline{\Phi}_W^C(\rho) = \tr_{\mathbb{C}^{n}}  \overline{W} \rho \overline{W}^*
\ee
It follows that
\be
E(C) = S_{\min}(\Phi_W)  = S_{\min}(\overline{\Phi}_W)
\ee
Our next result gives a universal upper bound for the minimum entropy of any product channel
of the form $\Phi \ot \overline{\Phi}$, depending only on the dimensions of the spaces
(a similar bound was derived in \cite{BranHoro09} for the case $s=n$).

\begin{thm}\label{thm:prod-bound}
Let $p = s/dn$, and assume that $sd/n \ge 1$, then
\be\label{ent-prod}
S_{\min}(\Phi \ot \overline{\Phi}) \le (1- p) \, \log (d^2 -1) - p \log p - (1-p) \log (1-p)
\ee
\end{thm}

\medskip
Theorem \ref{thm:prod-bound} will be proved in the Appendix.
We will now use Theorems \ref{thm1} and \ref{thm:prod-bound} to
demonstrate the existence of channels of the form $\Phi \ot \overline{\Phi}$ violating additivity.
For such a product channel the violation of additivity is given by
\be
\Delta S(\Phi) = S_{\min}(\Phi) + S_{\min}(\overline{\Phi}) - S_{\min}(\Phi \ot \overline{\Phi}) =
2 S_{\min}(\Phi) - S_{\min}(\Phi \ot \overline{\Phi})
\ee
Theorem \ref{thm1} guarantees the existence of subspaces satisfying the bound (\ref{thm:1}),
and hence also the existence of channels $\Phi$ for which  $S_{\min}(\Phi)$ satisfies the same bound.
Taking $s=n$ (so $r_1=r_2=1$), this implies the existence of channels for which
\be\label{ineq1.5}
S_{\min}(\Phi) > \log d - \frac{h_d(1,1)}{d}
\ee 
for sufficiently large $n$.
Combining the bounds (\ref{ineq1.5}) and (\ref{ent-prod}), and estimating
$\log (d^2 -1) \le 2 \log d$ in (\ref{ent-prod}), we obtain
\be\label{ineq2}
\Delta S(\Phi)  \ge p \, \log (p d^2) + (1-p) \, \log (1-p) - \frac{2}{d} \, h_d(1,1)
\ee
where $p = s/nd=1/d$.
Using (\ref{ineq2}) and the inequality $(1-p) \, \log (1-p) \ge - p$, we get for sufficiently large $n$
\be\label{ineq3}
\Delta S(\Phi)  \ge \frac{1}{d} \, \bigg[\log d  - 2 h_d(1,1) - 1 \bigg]
\ee
For $d > \exp [2 h_d(1,1) + 1]$  the right side of (\ref{ineq3}) is positive and hence these channels
violate additivity (recall that $h_d(1,1)$ is upper bounded uniformly in $d$). 
Furthermore, the method of proof shows that this violation occurs with positive probability for a randomly
selected subspace, and hence for a randomly selected channel.
Since the unital channels have measure zero in the set of all channels of fixed dimensions $s,n,d$, 
this implies the existence of non-unital channels which violate additivity.

\section{The Hastings bound}
This section contains an `abstract' version of the Hastings bound. Much of the notation was introduced
previously in \cite{FKM09}, and we will use several technical results from that paper.

\subsection{Notation}
${\cal M}_n$ will denote the algebra of complex $n \times n$ matrices;
the identity matrix will be written  $I$; ${\cal U}(n)$ will denote the group of unitary matrices.
The set of states in ${\cal M}_n$ is defined as
\be
{\cal S}_n = \{ \rho \in {\cal M}_n \,:\, \rho = \rho^* \ge 0, \,\, \tr \rho  =1 \}
\ee
The set of pure states in ${\cal M}_n$ is identified with the
unit vectors in $\mathbb{C}^n$ and denoted
\be
{\cal V}_n = \{ | \psi \ket   \in \mathbb{C}^n \,:\, \bra \psi | \psi \ket = 1 \}
\ee
We write ${\cal R}(s,n,d)$ for the set of all embeddings 
$W: \mathbb{C}^s \rightarrow \mathbb{C}^d \ot \mathbb{C}^n$, with $W^* W = I$.
There is a one-to-one correspondence between such embeddings 
and pairs of complementary channels
$\Phi_W \,:\, {\cal M}_s \rightarrow {\cal M}_n$, $\Phi_W^C \,:\, {\cal M}_s \rightarrow {\cal M}_d$
defined by
\be
\Phi_W(\rho) = \tr_{\mathbb{C}^d} W \rho W^*, \quad
\Phi_W^C(\rho) = \tr_{\mathbb{C}^n} W \rho W^*
\ee
Thus ${\cal R}(s,n,d)$ is also the set of all such pairs of conjugate channels.
The image of the pure input states under the action of a channel $\Phi^C \,:\, {\cal M}_s \rightarrow {\cal M}_d$ will be denoted
\be
{\rm Im}(\Phi^C) = \{ \Phi^C(| \phi \kb \phi |) \in {\cal S}_d \,:\, | \phi \ket \in {\cal V}_s \}
\ee

\subsection{Random embeddings}
We define a probability measure ${\cal P}_{s,n,d}$ on the set of embeddings ${\cal R}(s,n,d)$ as follows.
Let $W_0$ be a fixed embedding $W_0: \mathbb{C}^s \rightarrow \mathbb{C}^d \ot \mathbb{C}^n$
satisfying $W_0^* W_0 = I$. Then every embedding $W \in {\cal R}(s,n,d)$ can be written as
\be
W = U W_0, \quad U^* U = I
\ee
for some unitary matrix $U \in {\cal U}(nd)$. Let ${\rm Stab}(W_0)$ be the subgroup of unitary matrices 
which leave invariant every vector in the image of the embedding $W_0$. Then two unitary matrices
$U_1,U_2$ define the same embedding if $U_1^{-1} U_2 \in {\rm Stab}(W_0)$.
Thus ${\cal R}(s,n,d)$ can be identified with the left cosets
of the group of unitary
matrices with respect to the subgroup ${\rm Stab}(W_0)$.
Let $\Pi$ be the projection  from ${\cal U}(nd)$ onto these cosets.
Then the normalized Haar measure $Haar$ on ${\cal U}(nd)$ descends to
a normalized measure $\Pi^*(Haar)$ on this set of cosets, and this defines
our probability measure ${\cal P}_{s,n,d}$ on ${\cal R}(s,n,d)$.

\subsection{Definition of the tube}
We recall the notion of the `tube' at a state $\rho$, as defined in \cite{FKM09}.
First, for any $\rho \in  {\cal S}_d$ and $0 < \gamma < 1$ define $L_{\gamma}(\rho)$ to be 
the following line segment pointing from $\rho$ toward the maximally mixed state
$I/d$:
\be\label{def:L}
L_{\gamma}(\rho) = \Big\{ r \rho + (1-r) \frac{1}{d} I \,:\, \gamma \le r \le 1 \Big\}
\ee
Then the tube at $\rho$ is defined to be the set of states which
lie within a small distance of the set $L_{\gamma}(\rho)$. Also for any event $C \subset {\cal S}_d$, the tube at $C$ is the union
of the tubes at all states in $C$.

\begin{defn}
Let $\rho \in {\cal S}_d$, then the {\em $\tu$ at $\rho$} is defined as
\be\label{def:tube}
\tu(\rho) = \bigg\{ \theta \in {\cal S}_d \,:\, dist(\theta,L_{\gamma}(\rho)) \le 
2 \, \sqrt{\frac{\log n}{n}} +
13 \,d\, \sqrt{\frac{\log d}{s}} \bigg\}
\ee
where $dist(\theta,L(\rho))=
\inf_{\tau \in L(\rho)} \| \theta - \tau \|_{\infty}$.
For any output event $C \subset {\cal S}_d$ the {\em $\tu$ at $C$} is defined as
\be\label{def:tubeC}
\tu(C) = \bigcup_{\rho \in C} \, \tu(\rho)
\ee
\end{defn}

\subsection{Statement of the bound}
Suppose that for each triplet $(s,n,d)$ there is given an event
$C(s,n,d) \subset {\cal S}_d$. We want to find conditions which will show that for sufficiently large dimensions $s,n$
there is a nonzero probability that for a randomly selected embedding $W$
the event $C(s,n,d)$ will not contain any output states of the form $\Phi_W^C(| \phi \kb \phi |)$.
We analyze this by considering the sequence of complementary events, namely the events
$\{W \, : \, {\rm Im}(\Phi_W^C) \cap C(s,n,d) \neq \emptyset\}$, and showing that their probabilities approach zero
as $s,n \rightarrow \infty$.

\begin{thm}[The Hastings Bound]\label{prop-Hast}
Let $\{C(s,n,d) \subset {\cal S}_d \}$ be a collection of output events defined for each triplet of dimensions
$(s,n,d)$.
Fix $d \ge 2$, consider sequences $\{s_k, n_k\} \rightarrow \infty$, and define
\be
B_k = \{W \in {\cal R}(s_k,n_k,d) \,:\, {\rm Im}(\Phi_W^C) \cap C(s_k,n_k,d) \neq \emptyset \}
\ee
Suppose there is $\gamma \in (0,1)$ such that
\be\label{Hast-cond1}
\lim_{k \rightarrow \infty} \Big(d^2 \log n_k + n_k d \log d + (n_k - d) M(\gamma,k) - s_k \log (1 - \gamma)\Big) = - \infty
\ee
where
\be\label{def:Mk}
M(\gamma,k) = \sup \, \{ \tr \log \rho \,:\, \rho \in \tu(C(s_k,n_k,d)) \}
\ee
Then
\be
\lim_{k \rightarrow \infty} {\cal P}_{s_k,n_k,d}(B_k) = 0
\ee
\end{thm}

\medskip
As a consequence of the Theorem, if the conditions are satisfied then for
sufficiently large $k$ we have ${\cal P}_{s_k,n_k,d}(B_k) < 1$, and hence there must exist embeddings $W$ such that
$\Phi_W^C(| \phi \kb \phi |) \notin C(s_k,n_k,d)$ for any input state $| \phi \ket$.
Theorem \ref{prop-Hast} will be proved later in Section 5.
First we use it to deduce Theorems \ref{thm1} and \ref{thm2}.

\section{Proof of Theorem \ref{thm1}}
In this section we apply the Hastings bound to prove Theorem \ref{thm1}. Fix dimension $d$ and the parameter $h$.
In the following, we consider a sequence of integers $n$ large enough so that
we can choose $s=s(n)$ satisfying $r_1 \leq s/n \leq r_2$ for each $n$. We will prove the existence
of an integer $n_0$ such that (\ref{thm:1}) holds for all $n \ge n_0$, where
$n_0$ will not depend on the choice of $s(n)$.
Define
\be
C(s,n,d) = \bigg\{ \rho \in {\cal S}_d \,:\, S(\rho) \le \log d - h \Big(\frac{s}{nd}\Big) \bigg\}
\ee
Let $\lambda_i$ be the eigenvalues of $\rho$, then
\be
S(\rho) - \log d = - \frac{1}{d} \, \sum_{i=1}^d (\lambda_i d) \, \log (\lambda_i d)
\ee
Define
\be
f(x) = x \, \log x - x + 1
\ee
then it follows that
\be
C(s,n,d) = \bigg\{ \rho \in {\cal S}_d \,:\, \sum_{i=1}^d f(\lambda_i d) \ge h \Big(\frac{s}{n}\Big) \bigg\}
\ee
Next define
\be
F(x) = - \log x + x - 1
\ee
and note that for any $\theta \in {\cal S}_d$
\be
d \log d + \tr \log \theta = - \sum_{i=1}^d F(\theta_i d)
\ee
where $\{\theta_i\}$ are the eigenvalues of $\theta$. Thus recalling the definition (\ref{def:Mk}) it follows that
\be
d \log d +  M(\gamma,n) = - \inf \, \Big\{\sum_{i=1}^d F(\theta_i d) \,:\, 
\theta \in \tu(C(s,n,d)) \Big\}
\ee

\medskip
\noindent Now from definitions (\ref{def:tube}), (\ref{def:tubeC})  it follows that if
$\theta \in \tu(C(s,n,d))$  then for some $r \in [\gamma,1]$
\be
\theta_i = z_i + \epsilon_i, \quad
z_i = r \, \lambda_i + (1-r) \, \frac{1}{d}, \quad \sum_{i=1}^d \epsilon_i = 0
\ee
where the eigenvalues $\lambda_i$ satisfy
\be\label{f-bd-1}
\sum_{i=1}^d f(\lambda_i d) \ge h \Big(\frac{s}{n}\Big)
\ee
and where
\be
| \epsilon_i | \le 2 \, \sqrt{\frac{\log n}{n}} +
13 \,d\, \sqrt{\frac{\log d}{s}}
\ee
The Fannes inequality  \cite{Fannes73}, \cite{Audenaert07},  \cite{FKM09} implies that
\be\label{f-bd-2}
\Big| \sum_{i=1}^d f(\theta_i d) - \sum_{i=1}^d f(z_i d) \Big| \le \eta \equiv
d \, \epsilon_{m} \, ( \log d + \log \frac{1}{\epsilon_m} )
\ee
where
\be
\epsilon_{m} = \sum_{i=1}^d | \epsilon_i | \le  2 \,d\, \sqrt{\frac{\log n}{n}} +
13 \,d^2\, \sqrt{\frac{\log d}{s}}
\ee
Note that $\eta \rightarrow 0$ as $n,s \rightarrow \infty$.

\medskip
We now apply Lemma 12 from \cite{FKM09} which says that for all
$x >0$ and all $r \in [\gamma,1]$
\be
f(x) \le \frac{f(r x + 1 - r)}{f(1 - \gamma)}
\ee
Applying this to (\ref{f-bd-1}), (\ref{f-bd-2}) we deduce that
\be\label{f-bd-3}
\sum_{i=1}^d f(\theta_i d) \ge h \, f(1-\gamma) \, \Big(\frac{s}{n}\Big) - \eta
\ee
Thus we finally arrive at the inequality (putting $x_i = \theta_i d$)
\be\label{f-bd-4}
d \log d +  M(\gamma,n) \le - \inf_{\sum_{i=1}^d x_i = d} \, \bigg\{\sum_{i=1}^d F(x_i) \,:\, 
\sum_{i=1}^d f(x_i) \ge h \, f(1-\gamma) \, \Big(\frac{s}{n}\Big) - \eta \bigg\}
\ee
Define
\be
m_d(y) = \inf_{\sum_{i=1}^d x_i = d} \, \Big\{\sum_{i=1}^d F(x_i) \,:\, 
\sum_{i=1}^d f(x_i) \ge y \Big\}
\ee
then (\ref{f-bd-4}) can be written
\be\label{f-bd-41}
d \log d +  M(\gamma,n) \le - m_d\left(
h \, f(1-\gamma) \, \Big(\frac{s}{n}\Big) - \eta \right)
\ee
In Section 5.7 of \cite{FKM09} the following identity was derived:
\be
m_d(y) =  \inf_{z > 1} \Big\{ - \log z - (d-1) \log \frac{d-z}{d-1} \,:\, z \log z + (d-z) \log \frac{d-z}{d-1} = y \Big\}
\ee
It was also shown in \cite{FKM09} that the function $m_d$ is increasing and hence has an inverse
$m_d^{-1}$. Given $y > 0$ there is a unique $z > 1$ satisfying $z \log z + (d-z) \log \frac{d-z}{d-1} = y$,
and this function also has an inverse. Thus
\be
m_d^{-1}(w) = z \log z + (d-z) \log \frac{d-z}{d-1}
\ee
where $z$ is the unique solution of $- \log z - (d-1) \log \frac{d-z}{d-1} = w$. Since both of these functions
are increasing this can be written as the minimization
\be
m_d^{-1}(w) = \inf_{z > 1} \Big\{ z \log z + (d-z) \log \frac{d-z}{d-1} \,:\, 
- \log z - (d-1) \log \frac{d-z}{d-1} = w \Big\}
\ee
Thus recalling (\ref{def:Q}) we have 
\be
h_d(x,y) = \inf_{0 < \gamma < 1} \, \frac{1}{x \, f(1 - \gamma)} \, m_d^{-1}\Big(- \log (1 - \gamma) \, y\Big)
\ee
Let $\gamma_m$ be the value where the infimum is achieved, then
\be
m_d(x \, f(1 - \gamma_m) \, h_d(x,y)) = - \log (1 - \gamma_m) \, y
\ee

Returning to (\ref{f-bd-41}), and using the bound $s/n \ge r_1$,
\be\label{f-bd-5}
d \log d +  M(\gamma,n) \le  - m_d\left(h \, f(1-\gamma) \, r_1 - \eta \right)
\ee
By assumption $h > h_d(r_1,r_2)$, and also $\eta \rightarrow 0$ as $n \rightarrow \infty$, hence
there is $\delta > 0$ such that for $n$ sufficiently large
\be
h \, f(1-\gamma) \, r_1 - \eta > h_d(r_1,r_2) \, f(1-\gamma) \, r_1 + \delta
\ee
and thus
\be\label{f-bd-6}
d \log d +  M(\gamma,n) \le - m_d\left(h_d(r_1,r_2) \, f(1-\gamma) \, r_1 + \delta\right)
\ee
Furthermore as was shown in \cite{FKM09} 
\be
m_d'(y) = \frac{d(1 - z^{-1})}{d \log z  - y} > \frac{1}{z}
\ee
where $z$ is the unique solution of $z \log z + (d-z) \log \frac{d-z}{d-1} = y$. The maximum
value of $z$ is $d$ hence we obtain
\be
m_d'(y) \ge \frac{1}{d}
\ee
Thus from (\ref{f-bd-6}) (applying the mean value theorem)
\be\label{f-bd-7}
d \log d +  M(\gamma,n) \le - m_d(h_d(r_1,r_2) \, f(1-\gamma) \, r_1) - \frac{1}{d} \, \delta
\ee
Setting $\gamma = \gamma_m$ we have
\be
m_d(h_d(r_1,r_2) \, f(1-\gamma_m) \, r_1) = - \log (1 - \gamma_m) \, r_2
\ee
Thus finally returning to (\ref{Hast-cond1}) we have
\be\label{f-bd-8}
& & d^2 \log n + n d \log d + (n - d) M(\gamma_m,n) - s \log (1 - \gamma_m) \nonumber \\
 &=&
 d^2 (\log n + \log d)- \frac{sd}{n} \log (1 - \gamma_m)
+(n-d)\bigg(d \log d +  M(\gamma_m,n) 
-\frac{s}{n}\log (1 - \gamma_m)
\bigg) \nonumber \\
 &\le &    d^2 (\log n + \log d)- r_2 d \log (1 - \gamma_m) 
+(n-d)\bigg(\log (1-\gamma_m) \,r_2 -\frac{1}{d} \,\delta
-\frac{s}{n}\log (1 - \gamma_m)
\bigg) \nonumber \\
&\le &     
d^2 (\log n + \log d)- r_2 d \log (1 - \gamma_m) - (n-d) \, \frac{1}{d} \, \delta 
\ee
where we used $s/n \le r_2$. Since $\delta > 0$ the right side of (\ref{f-bd-8}) diverges to $- \infty$
as $n \rightarrow \infty$. Thus applying Theorem \ref{prop-Hast} yields the result.

\section{Proof of Theorem \ref{thm2}}
Following the steps of the proof of Theorem \ref{thm1} leads to
\be\label{f-bd-4a}
d \log d +  M(\gamma,k) \le - m_d\left(
h \, f(1-\gamma) \, \Big(\frac{s_k}{n_k}\Big) - \eta \right)
\ee
where
\be
\eta = d \, \epsilon_{m} \, ( \log d + \log \frac{1}{\epsilon_m} ), \quad
\epsilon_{m} \le  2 \,d\, \sqrt{\frac{\log n_k}{n_k}} +
13 \,d^2\, \sqrt{\frac{\log d}{s_k}}
\ee
The assumptions that $s_k \rightarrow \infty$ and $n_k \log s_k/s_k^{3/2} \rightarrow 0$ imply that 
\be\label{eta-bd}
\frac{\eta}{s_k/n_k} \rightarrow 0 \quad
\mbox{as $k \rightarrow \infty$}
\ee
and hence the first term $h \, f(1-\gamma) \, (s_k/n_k)$ on the right side of (\ref{f-bd-4a}) dominates $\eta$.
Since $s_k/n_k \rightarrow 0$ we must consider the behavior of $m_d(y)$ as
$y \rightarrow 0$.

\begin{lemma}\label{lem:md->0}
There is $y_0 > 0$ such that $y^{-1} m_d(y)$ is decreasing for all
$0 < y \le y_0$. Furthermore
\be\label{lem:md->1}
\lim_{y \rightarrow 0} \frac{m_d(y)}{y} = 1
\ee
\end{lemma}

Lemma \ref{lem:md->0} will be proved in the Appendix.
We now use it to analyze (\ref{f-bd-4a}). Since $s_k/n_k \rightarrow 0$, and using
(\ref{eta-bd}), it follows from (\ref{lem:md->1}) that for any $\epsilon > 0$ there is $k_0$ such that for 
$k > k_0$,
\be
m_d\left(
h \, f(1-\gamma) \, \Big(\frac{s_k}{n_k}\Big) - \eta \right) \ge (1 - \epsilon) \, h \, f(1-\gamma) \, \Big(\frac{s_k}{n_k}\Big) - \eta
\ee
Hence from (\ref{f-bd-4a})
\be\label{f-bd-5a}
d \log d +  M(\gamma,k) \le - (1 - \epsilon) \,
h \, f(1-\gamma) \, \Big(\frac{s_k}{n_k}\Big) + \eta
\ee
Turning now to (\ref{Hast-cond1}) we have for $k > k_0$
\begin{align}\label{f-bd-8a}
d^2 \log n_k 
& + n_k d \log d 
+ (n_k - d) M(\gamma,k) - s_k \log (1 - \gamma)  \nonumber \\
& \le     d^2 (\log n_k + \log d)- \frac{s_kd}{n_k} \log (1 - \gamma)
\nonumber \\
&\qquad \qquad -(n_k-d)\bigg((1 - \epsilon) \,
h \, f(1-\gamma) \, \Big(\frac{s_k}{n_k}\Big) -\eta
+\frac{s_k}{n_k}\log (1 - \gamma)
\bigg) \nonumber \\
&  = 
d^2 (\log n_k + \log d)- \frac{s_kd}{n_k} \log (1 - \gamma) \nonumber \\
& \qquad \qquad - \frac{s_k(n_k-d)}{n_k}
\bigg((1 - \epsilon) \,
h \, f(1-\gamma) \, -\eta\Big(\frac{n_k}{s_k}\Big) 
+\log (1 - \gamma)
\bigg)
\end{align}
By assumption there is $\gamma_m$ such that
\be
h > \frac{- \log (1 - \gamma_m)}{f(1-\gamma_m)}
\ee
From (\ref{eta-bd}) it follows that there is $\delta > 0$ such that for $k$ sufficiently large
\be
(1 - \epsilon) \,
h \, f(1-\gamma_m) \,  - \eta \, \frac{n_k}{s_k} +  \log (1 - \gamma_m) > \delta
\ee
Thus from (\ref{f-bd-8a})
\be\label{f-bd-9a}
&d^2 \log n_k + n_k d \log d + (n_k - d) M(\gamma_m,k) - s_k \log (1 - \gamma_m) & \nonumber \\
&&  \hskip-2in \le
d^2 (\log n_k + \log d)- \frac{s_kd}{n_k} \log (1 - \gamma_m) 
- \frac{s_k(n_k-d)\,\delta}{n_k}
\ee
Since $s_k/\log n_k \rightarrow \infty$ the right side of (\ref{f-bd-9a}) diverges to $- \infty$
as $k \rightarrow \infty$.
\qed

\section{Proof of the Hastings bound}
First we recall some of the ideas and notation from \cite{FKM09}.
The set of eigenvalues of a state $\rho$ is denoted ${\rm spec}(\rho)$.
Also ${\Delta}_d$ denotes the simplex of $d$-dimensional probability distributions:
\be\label{def:Delta}
{\Delta}_d = \{(x_1,\dots,x_d) \subset \mathbb{R}^d \,:\, x_i \ge 0, \,\, \sum_{i=1}^d x_i = 1 \}
\ee

\subsection{Random states}
The pure states ${\cal V}_n$ can be identified with the unit sphere in $\mathbb{R}^{2n}$.
This provides a probability measure on ${\cal V}_n$, namely the normalized uniform measure
which we denote $\sigma_n$. Thus saying that $| \psi \ket \in {\cal V}_n$
is a random vector means that $| \psi \ket$ has the uniform distribution $\sigma_n$.

\medskip
Let $| z \ket = \bmx z_1 & \cdots & z_{dn} \emx^T$ be a unit vector in ${\cal V}_{d n}$. Then $|z\ket$ can be written as a $n \times d$ matrix $M$, with entries
\be
M_{ij}(z) = z_{(i-1)d + j}, \quad i=1,\dots n, \,\, j=1,\dots,d
\ee
satisfying $\tr M^* M = \sum_{ij} |z_{ij}|^2 = 1$.
Define the map $G \,:\, {\cal V}_{d n} \rightarrow {\cal M}_d$ by
\be\label{def:G}
G(z) = M(z)^* M(z)
\ee
The eigenvalues of $G(z)$  lie in  ${\Delta}_d$.
When $|z\ket \in {\cal V}_{d n}$ is a random vector,  the probability density $\mu_{d,n}$ of these eigenvalues
is known explicitly \cite{LloydPagels}, \cite{ZycSom01}:
for any event $A \subset \Delta_d$
\be\label{mu-density}
\mu_{d,n}(A) = Z(n,d)^{-1} \, \int_{A} \prod_{1 \le i < j \le d} (w_i - w_j)^2 
\prod_{i=1}^d w_i^{n-d} \, \delta\bigg(\sum_{i=1}^d w_i - 1\bigg) \, [dw]
\ee
where $Z(n,d)$ is a normalization factor. 
We recall the following bound which was derived in \cite{FKM09}.
\begin{lemma}\label{lem:bound-mu}
For all $d$, for $n$ sufficiently large, and for any event $A \subset {\Delta}_d$,
\be\label{mu3a}
\mu_{d,n}(A) 
\le  \frac{1}{(d-1)!} \, \exp \Big[d^2 \log n + (n-d) \, d \, \log d + (n-d) \, \sup_{w \in A} \sum_{i=1}^d \log w_i \Big]
\ee
\end{lemma}

Now let $C \subset {\cal S}_d$ be any set 
which is invariant under conjugation by
every unitary matrix in ${\cal U}(d)$. Then the event $\{| \psi \ket \,:\, \tr_2 | \psi \kb \psi | \in C \}$ 
depends only on the eigenvalues of $\tr_2 | \psi \kb \psi |$, and thus its probability is determined by
$\mu_{d,n}$. For an arbitrary set $C \subset {\cal S}_d$ 
we define
\be
{\tilde C} = \{ V \, \rho \, V^* \,:\, \rho \in C, \quad V \in {\cal U}(d) \}, \quad
{\rm spec}(C) = \bigcup_{\rho \in C} \, {\rm spec}(\rho)
\ee
Then ${\tilde C}$ is invariant under conjugation by an arbitrary unitary matrix,
and ${\rm spec}(C) = {\rm spec}({\tilde C})$.
Hence from (\ref{mu3a}) we deduce the bound
\be\label{bd-G*-1}
G^{*}(\sigma_{nd})(C) &\le& G^{*}(\sigma_{nd})({\tilde C}) \nonumber \\
& = & \mu_{d,n}({\rm spec}(C)) \nonumber \\
& \le &  \frac{1}{(d-1)!} \, \exp \Big[d^2 \log n + (n-d) \, d \, \log d + 
(n-d) \, \sup_{w \in {\rm spec}(C)} \sum_{i=1}^d \log w_i \Big] \nonumber \\
& = & \frac{1}{(d-1)!} \, \exp \Big[d^2 \log n + (n-d) \, d \, \log d + 
(n-d) \, \sup_{\rho \in C} \tr \log \rho \Big]
\ee

\subsection{Random embeddings yield random output states}
Let $W \in {\cal R}(s,n,d)$ be a random embedding. Then
for any pure state $| \phi \ket \in {\cal V}_s$ the vector $| \psi \ket = W | \phi \ket = U W_0 | \phi \ket$ is a
uniform random vector in $\mathbb{C}^{nd}$. Thus $\Phi_W^C(| \phi \kb \phi|) = \tr_2 | \psi \kb \psi |$ is the reduced
density matrix of a random vector. This remains true if $| \phi \ket$ is a random input state.
To formalize this relation we define the map
\be\label{def:H}
H: {\cal R}(s,n,d) \times {\cal V}_s \rightarrow {\cal M}_d, \quad
(W, | \phi \ket) \mapsto \Phi_W^C(| \phi \kb \phi|)
\ee

\medskip
\begin{lemma}\label{lem1}
\be
H^*({\cal P}_{s,n,d} \times \sigma_s) = G^*(\sigma_{d n})
\ee
\end{lemma}
The proof of
Lemma \ref{lem1} is very similar to the proof of Lemma 7 in \cite{FKM09} and so we omit the details here.
Lemma \ref{lem1} implies that if $W$ is chosen randomly according to
the measure ${\cal P}_{s,n,d}$ and $| \phi \ket$ is chosen randomly and uniformly in ${\cal V}_s$, then
the eigenvalues of the matrix
$\Phi_W^C(| \phi \kb \phi|)$ will have the distribution $\mu_{d,n}$, as defined above in (\ref{mu-density}).

\subsection{Typical channels}
For a random embedding $W$ `most' output states
of the channel $\Phi_W^C$
are close to the maximally mixed state. More precisely,
an embedding $W$ will be called typical if $\Phi_W^C$
maps at least one half of input states into a small ball centered at the maximally
mixed output state $I/d$. The ball is defined as follows:
\be\label{def:B-n}
B_d(n) = \bigg\{ \rho \in {\cal S}_d \,:\, \Big\| \rho - \frac{1}{d} I \Big\|_{\infty} \le 2 \sqrt{\frac{\log n}{n}} \bigg\}
\ee

\begin{defn}
An embedding $W$ is called {\em typical} if with probability at least $1/2$ a randomly
chosen input state is mapped by $\Phi_W^C$ into the set $B_d(n)$. The set of typical embeddings is denoted $T$:
\be
T = \bigg\{ W \,:\, \sigma_s\Big(| \phi \ket \,:\, \Phi_W^C(| \phi \kb \phi |) \in B_d(n)\Big) \ge 1/2 \bigg\}
\ee
\end{defn}

As the next result shows, for large $n$ most embeddings are typical. This result was proved in
\cite{FKM09} and we just quote the result here (note that in \cite{FKM09} the definition of $B_d(n)$
contained a free parameter $b$ which was required to be at least $\sqrt{3}$ -- here we have set $b=2$).

\begin{lemma}\label{lem:prob-Tc}
For 
each $d\geq 2$ taking $n$ sufficiently large,
and for all $s$, 
\be\label{lem:prob-Tc1}
{\cal P}_{s,n,d}(T^c) \le \frac{2 \, d}{(d-1)!} \,\, \exp[-\alpha \, d^2 \, \log n], \quad
\alpha =  \frac{4(n-d)}{3n} - 1
\ee
\end{lemma}

Thus as $n \rightarrow \infty$ with high probability a randomly chosen embedding will  lie in the set $T$.
In particular ${\cal P}_{s,n,d}(T^c) < 1$ for $n$ sufficiently large. 

\medskip
The next result says that for a typical embedding $W$ there is a fixed fraction of input states
which are mapped by $\Phi_W^C$ into the tube at any output state $\rho$. This result is crucial
for the proof and differs in some significant ways from the related proof in \cite{FKM09},
thus we include full details in the Appendix.

\begin{lemma}\label{lem:uniform-lwr-bnd}
Let $d,s \ge 2$, then for $n$ sufficiently large, for all
$W \in T$ and $\rho \in {\rm Im}(\Phi_W^C)$
\be
\sigma_s\Big(| \phi \ket \,:\, \Phi_W^C(| \phi \kb \phi |) \in \tu(\rho)\Big) \ge \frac{1}{4} \, \bigg( 1-\gamma \bigg)^{s-1}
\ee
\end{lemma}

\subsection{The proof}
Define
\be
E_k = \{ (W, | \phi \ket) \,:\, W \in B_k, \,\, \Phi_W^C(| \phi \kb \phi |) \in \tu(C(s_k,n_k,d)) \}
\ee
The proof will proceed by proving upper and lower bounds for 
the probability of $E_k$, that is $({\cal P}_{s,n,d} \times \sigma_s)(E_k)$.

\medskip
For the upper bound, note that by Lemma \ref{lem1},
\be\label{uppr1}
({\cal P}_{s,n,d} \times \sigma_s)(E_k) & \le &
({\cal P}_{s,n,d} \times \sigma_s)\{ (W, | \phi \ket) \,:\,  \Phi_W^C(| \phi \kb \phi |) \in \tu(C(s_k,n_k,d)) \} \nonumber \\
& = &
({\cal P}_{s,n,d} \times \sigma_s)(H^{-1}(\tu(C(s_k,n_k,d)))) \nonumber \\
&=& H^*({\cal P}_{s,n,d} \times \sigma_s)(\tu(C(s_k,n_k,d))) \nonumber \\
&=&
G^*(\sigma_{d n})(\tu(C(s_k,n_k,d)))
\ee
Recall the definition of $M(\gamma,k)$ in (\ref{def:Mk}).
Using the bound (\ref{bd-G*-1}) we deduce
\be\label{upper4}
({\cal P}_{s,n,d} \times \sigma_s)(E_k) 
 & \le & \frac{1}{(d-1)!} \exp \, \Big[d^2 \log n_k + (n_k -d) d \log d + (n_k -d) M(\gamma,k) \Big] \nonumber \\
 & = & \alpha(d) \, \exp \, \Big[d^2 \log n_k + n_k d \log d + (n_k - d) M(\gamma,k) \Big]
\ee
where $\alpha(d) = \exp [-d^2 \log d]/(d-1)!$ .

\medskip
The derivation of the lower bound is very similar to that in \cite{FKM09}, however we include it
here for completeness. First we write
\bee
({\cal P}_{s,n,d} \times \sigma_s)(E_k) &=& \mathbb{E}_{W}[1_{B_k} \, \sigma_s(| \phi \ket \,:\, \Phi_W^C(| \phi \kb \phi |) \in \tu(C(s_k,n_k,d)))] \\
& \ge & \mathbb{E}_{W}[1_{B_k \cap T} \, \sigma_s(| \phi \ket \,:\, \Phi_W^C(| \phi \kb \phi |) \in \tu(C(s_k,n_k,d)))]
\eee
where $\mathbb{E}_{W}$ denotes expectation over ${\cal R}(s_k,n_k,d)$ with respect to the 
measure ${\cal P}_{s,n,d}$, and $1_{B_k \cap T}$ is the characteristic function of the event
$B_k \cap T$.
Given that $W \in B_k$ there is a state $| \theta \ket \in \mathbb{C}^s$ such that
\be
\Phi_W^C(| \theta \kb \theta |) \in C(s_k,n_k,d)
\ee
Since $\tu(\Phi_W^C(| \theta \kb \theta |)) \subset \tu(C(s_k,n_k,d))$ it follows that
\be\label{ineq3a}
({\cal P}_{s,n,d} \times \sigma_s)(E_k) \ge 
\mathbb{E}_{W}[1_{B_k \cap T} \, \sigma_s(| \phi \ket \,:\, \Phi_W^C(| \phi \kb \phi |) \in \tu(\Phi_W^C(| \theta \kb \theta |)))]
\ee
Applying Lemma \ref{lem:uniform-lwr-bnd} to (\ref{ineq3a}) gives
\be\label{ineq3b}
({\cal P}_{s,n,d} \times \sigma_s)(E_k) & \ge & \frac{1}{4} \, \bigg( 1 - \gamma \bigg)^{s_k-1} \,
\mathbb{E}_{W}[1_{B_k \cap T}] \\ 
& = & \frac{1}{4} \,\bigg( 1 - \gamma \bigg)^{s_k-1} \, {\cal P}_{s,n,d}(B_k \cap T) \\
& \ge & \frac{1}{4} \,\bigg( 1 - \gamma \bigg)^{s_k-1} \, ({\cal P}_{s,n,d}(B_k) - {\cal P}_{s,n,d}(T^c)) 
\ee

\medskip
Putting together the upper and lower bounds for $({\cal P}_{s,n,d} \times \sigma_s)(E_k)$
produces the following bound: for all $d \ge 2$,
for all $0 < \gamma < 1$,  and for $n$ sufficiently large
\be\label{ineq:16}
{\cal P}_{s,n,d}(B_k) & -& {\cal P}_{s,n,d}(T^c) \nonumber \\
& \le &  
4 \, \bigg( \frac{1}{1 - \gamma} \bigg)^{s_k-1} \, 
({\cal P}_{s,n,d} \times \sigma_s)(E_k)  \\
& \le &   4 \alpha(d) \,\bigg( \frac{1}{1 - \gamma} \bigg)^{s_k-1} \, 
\exp \, \Big[d^2 \log n_k + n_k d \log d + (n_k - d) M(\gamma,k) \Big] \nonumber \\
&= &  4 \alpha(d) \, (1-\gamma) \,
\exp\left[d^2 \log n_k + n_k d \log d + (n_k - d) M(\gamma,k) - s_k \log (1-\gamma)\right] \nonumber
\ee
Note that Lemma \ref{lem:prob-Tc}  implies ${\cal P}_{s,n,d}(T^c) \rightarrow 0$ as $k \rightarrow \infty$. Also,
by assumption there is $\gamma$ such that
\be
d^2 \log n_k + n_k d \log d + (n_k -d) M(\gamma,k) - s_k \log (1-\gamma) \rightarrow - \infty
\ee
as $k \rightarrow \infty$. By choosing this value for $\gamma$ we deduce that
${\cal P}_{s,n,d}(B_k) \rightarrow 0$ as required.
\qed

\medskip
\bigskip

\noindent{\bf Acknowledgments:}
This collaboration began
at ``the University of Arizona FRG Workshop'' in June 2009
(the grant number: DMS-0757581),
and a lot of progress was made 
and the results were presented
during 
the July 2009 workshop
``Thematic Program on Mathematics in Quantum Information''
at the Fields Institute.
The authors are grateful to the organizers of these workshops.

\pagebreak
\appendix

\section{Proof of Theorem \ref{thm:prod-bound}}
The bound for $S_{\min}(\Phi \ot \overline{\Phi})$ is obtained using a Kraus representation $\Phi(\rho) = \sum_{i=1}^d A_i \rho A_i^{*}$
and the maximally entangled state.
Let $|\hat{\psi}\rangle$ and $|\hat{\phi}\rangle$ be the maximally entangled states
on $\mathbb{C}^s \otimes \mathbb{C}^s$ and 
$\mathbb{C}^n\otimes \mathbb{C}^n$ respectively. Then
\begin{align}  
(\Phi \otimes \overline{\Phi}) (|\hat{\psi}\rangle\langle\hat{\psi}| )
&= \sum_{i,j=1}^d  ( A_i \otimes \overline{A_j} )
|\hat{\psi}\rangle\langle\hat{\psi}| 
(A_i^\ast \otimes  A_j^T ) \\
&= \frac{n}{s}\sum_{i=1}^d  (A_i A_i^* \otimes I) 
|\hat{\phi}\rangle\langle\hat{\phi}|(A_i A_i^* \otimes I)
+ \sum_{i \not= j} (A_i \otimes \overline{A_j} )
|\hat{\psi}\rangle\langle\hat{\psi}| 
(A_i^\ast \otimes  A_j^T )
\end{align}
where we used the identity $(A \otimes \overline{A} )
|\hat{\psi}\rangle= \sqrt{n/s} \,
(A A^* \otimes I) 
|\hat{\phi}\rangle$.
Note that
$\sum_i A_i A_i^* = \Phi(I_s)$, therefore
\bee
\sum_{i=1}^d \tr A_i A_i^* = \tr \Phi(I_s) = \tr I_s = s
\eee
Also
\bee
\langle \hat{\phi}| (A_i A_i^* \otimes I) 
|\hat{\phi}\rangle
= \frac{1}{n} \tr [A_i A_i^*] 
\eee
hence
\bee
\langle \hat{\phi}|\sum_{i=1}^d  (A_i A_i^* \otimes I) 
|\hat{\phi}\rangle\langle\hat{\phi}|(A_i A_i^* \otimes I) |\hat{\phi} \rangle
= \frac{1}{n^2}\sum_{i=1}^d \left( \tr [A_i A_i^*] \right)^2
\geq \frac{s^2}{d n^2}
\eee
Hence
\be
\langle \hat{\phi}|  
(\Phi \otimes \overline{\Phi}) (|\hat{\psi}\rangle\langle\hat{\psi}| )
|\hat{\phi} \rangle
\geq \frac{s}{dn}
\ee

This shows that one of the eigenvalues of $(\Phi \otimes \overline{\Phi}) (|\hat{\psi}\rangle\langle\hat{\psi}| )$ 
is larger than or equal to $p = s/(dn)$. 
The rank of this matrix is $d^2$, hence it has at most $d^2-1$ other
nonzero eigenvalues.
Given  that $sd\geq n$, the entropy is maximized when these other eigenvalues are equal to
$(1-p)/(d^2-1)$.
This implies that the entropy cannot be larger than
\bee
S\left( (\Phi \otimes \overline{\Phi}) (|\hat{\psi}\rangle\langle\hat{\psi}| )\right)
\leq
g(p) = -p\log p - (1-p) \log \left(\frac{1-p}{d^2 -1}\right)
\eee

\section{Proof of Lemma \ref{lem:uniform-lwr-bnd}}
The result is very similar to the proof of Lemma 11 in \cite{FKM09},
but with important differences in detail.
Let $| \psi \ket$ be a fixed state in ${\cal V}_s$, and let $| \theta \ket$ be a random pure state
in ${\cal V}_s$, with probability distribution $\sigma_s$. 
We write
$x = \bra \psi | \theta \ket$, and let $| \phi \ket$ be the state orthogonal to
$| \psi \ket$ such that
\be\label{def:phi}
| \theta \ket = x \, | \psi \ket + \sqrt{1 - |x|^2} \, | \phi \ket
\ee
Thus $| \phi \ket$ is also a random state, defined by its relation to the uniformly random state
$| \theta \ket$ in (\ref{def:phi}). The following result was proved in \cite{FKM09}.

\begin{prop}\label{random-states}
$x$ and $| \phi \ket$ are independent. $| \phi \ket$ is a random vector
in ${\cal V}_{s-1}$ with distribution $\sigma_{s-1}$.
For all $0 \le t \le 1$
\be\label{Prop:eqn1}
\sigma_s \{ | \theta \ket \,:\, |\bra \psi | \theta \ket| = |x| > t \} = (1 - t^2)^{s-1}
\ee
\end{prop}

Proposition \ref{random-states} implies that
as $s \rightarrow \infty$ the overlap $x = \bra \psi | \theta \ket$ becomes concentrated around zero.
In other words, with high probability a randomly chosen state will be almost
orthogonal to any given fixed state. As a  consequence, from (\ref{def:phi}) it follows that
$| \phi \ket$ will be almost equal to $| \theta \ket$. This statement is made precise
by noting that
\be
\| | \theta \ket - | \phi \ket \|_{2} \leq \sqrt{2} \, |\bra \psi | \theta \ket|
\ee
Then (\ref{Prop:eqn1}) immediately implies that
\be\label{prop1a}
\sigma_s(| \theta \ket \,:\, \| | \theta \ket - | \phi \ket \|_{2} > t) 
\leq  \left(1 - \frac{t^2}{2} \right)^{s-1}
\ee

\medskip
The second property relies on the form of the conjugate channel $\Phi^C$. 
If the Kraus decomposition for $\Phi$ is
$\Phi(\rho) = \sum_{i=1}^d A_i \rho A_i^*$ then the Kraus decomposition for
$\Phi^C$ is
\be
\Phi^C(\rho) = \sum_{k,l=1}^d \tr (  A_k\rho A_l^{\ast}) \, | k \kb l |
\ee
For any fixed channel $\Phi$ and random state $| \theta \ket$, with high probability
the norm of the matrix $\Phi^C(| \theta \kb \psi |)$ is small,
and approaches zero as $n \rightarrow \infty$. We will prove the following bound:
for any $\Phi \in {\cal R}(s,n,d)$, and for all $0 \le t \le 1$,
\be\label{prop2}
\sigma_s\Big(| \theta \ket \,:\, \| \Phi^C(| \theta \kb \psi |) \|_2 > t\Big) 
\le d^2 \, \left(1 - \left(\frac{t}{d}\right)^2\right)^{s-1}
\ee

\medskip
As a first step toward deriving (\ref{prop2}), note that for any vectors
$| u \ket$ and $| v \ket$,
\be\label{prop2a}
\| \Phi^C(| u \kb v |) \|_2 = 
\left(\sum_{k,l =1}^d 
 |\bra v | A_l^\ast A_k |u \ket |^2
 \right)^{\frac{1}{2}}
\leq d  \, \max_{k,l} |\bra v | A_l^\ast A_k |u\ket|.
\ee
Since $\sum_{i=1}^d A_i^* A_i = I$ it follows that $\| A_i \|_{\infty} \le 1$ for all
$i=1,\dots,d$, which implies that 
\be\label{prop1b}
\| \Phi^C(| u \kb v |) \|_2 \le d \,\| | u \ket \|_{2} \, \| | v \ket \|_{2} 
\ee

To derive (\ref{prop2}) we apply (\ref{prop2a}) with $u = \theta$ and $v = \psi$ and deduce that
\be\label{prop2b}
\sigma_s\Big( | \theta \ket \,:\, \| \Phi^C(| \theta \kb \psi |) \|_2 > t\Big) & \le &
\sigma_s
\left( | \theta \ket \,
:\, \max_{k,l} |\bra \psi | A_l^\ast A_k | \theta \ket| > \frac{t}{d}\right) \nonumber \\
& \le & d^2 \,
\sigma_s \left( | \theta \ket \,
:\, |\bra \psi | A_l^\ast A_k | \theta \ket| > \frac{t}{d}\right) \nonumber \\
& \leq & d^2 \, \left(1 - \left(\frac{t}{d}\right)^2 \right)^{s-1}
\ee
where the last equality follows from (\ref{Prop:eqn1}).
Note that for each $k,l$, the above 
 $| A_k^\ast A_l \psi\ket$ is a fixed vector
with norm less than $1$.

\medskip
With these ingredients in place the proof of Lemma \ref{lem:uniform-lwr-bnd} can proceed.
By assumption $\Phi$ is a channel belonging to the typical set $T$, and
$\rho = \Phi^C(|\psi \kb \psi|)$ is some state in ${\rm Im}(\Phi^C)$.
Let $| \theta \ket$ be a random input state, then as in (\ref{def:phi}) we write
\be\label{def:phi2}
| \theta \ket = x \, | \psi \ket + \sqrt{1 - |x|^2} \, | \phi \ket \nonumber
\ee
It follows that
\be\label{tube2}
| \theta \kb \theta | = |x|^2 \, | \psi \kb \psi | + (1 - |x|^2) \, | \phi \kb \phi |
+ \sqrt{1 - |x|^2} \, (x \, | \psi \kb \phi | + \overline{x} \, | \phi \kb \psi |)
\ee
Write $r = |x|^2$, then (\ref{tube2}) yields
\be\label{tube3}
\Phi^C(|\theta\rangle\langle\theta|) 
&& \hskip-0.2in - \left(
r \Phi^C(|\psi\rangle\langle\psi|) + 
(1-r)\frac{1}{d}I\right) \nonumber \\
&&\quad = (1-r) \left( \Phi^C |\phi\rangle\langle\phi|)- \frac{1}{d}I\right)
+\sqrt{r(1-r)}
\Phi^C\left( e^{i \xi} \, |\psi\rangle\langle\phi| + e^{- i \xi} \, |\phi\rangle\langle\psi| \right)
\ee
where $\xi$ is the phase of $x$. Since $r \le 1$ this implies
\be\label{tube4}
\bigg\| \Phi^C(|\theta\rangle\langle\theta|) 
 - \left(
r\Phi^C(|\psi\rangle\langle\psi|) + 
(1-r)\frac{1}{d}I\right) \bigg\|_{\infty} 
 \le \bigg\|  \Phi^C (|\phi\rangle\langle\phi|)- \frac{1}{d}I \bigg\|_{\infty} +
\bigg\| \Phi^C (|\psi\rangle\langle\phi|) \bigg\|_{\infty}
\ee
Referring to the definition (\ref{def:tube}) of $\tu (\rho)$,
recall that $\Phi^C(|\theta\rangle\langle\theta|)$ belongs to  $\tu(\rho)$ if and only if for some $r$
satisfying $\gamma \le r \le 1$,
\be\label{tube5}
\bigg\| \Phi^C(|\theta\rangle\langle\theta|) 
- \left(
r\Phi^C(|\psi\rangle\langle\psi|) + 
(1-r)\frac{1}{d}I\right) \bigg\|_{\infty} \le 2\, \sqrt{\frac{\log n}{n}} + 13 \,d\, \sqrt{\frac{\log d}{s}}
\ee 
Define the following three events in ${\cal V}_s$:
\be\label{def:events}
A_1 & = & \{| \theta \ket \,:\, r = |\bra \psi | \theta \ket|^2 \ge \gamma \} \\
A_2 &= & \bigg\{| \theta \ket \,:\, \bigg\|  \Phi^C (|\phi\rangle\langle\phi|)- \frac{1}{d}I \bigg\|_{\infty}
\le    \sqrt{\frac{48 d^2 \log d}{s}} + 2 \, \sqrt{\frac{\log n}{n}}\bigg\} \\
A_3 &= & \bigg\{| \theta \ket \,:\, \bigg\| \Phi^C (|\psi\rangle\langle\phi|) \bigg\|_{\infty} \le 
\sqrt{\frac{6 d^2 \log d}{s}} + \sqrt{\frac{ 12 d^2\log d}{s}} \bigg\}
\ee
It follows from (\ref{tube4}) and (\ref{tube5}) that
\be
A_1 \cap A_2 \cap A_3 \subset \{| \theta \ket \,:\, 
\Phi^C(|\theta\rangle\langle\theta|) \in \tu(\rho) \}
\ee
Furthermore by Proposition \ref{random-states}, $A_1$ is independent of $A_2$ and $A_3$, hence
\be\label{tube6}
\sigma_s(\Phi^C(|\theta\rangle\langle\theta|) \in \tu(\rho)) \ge
\sigma_s(A_1 \cap A_2 \cap A_3) = 
\sigma_s(A_1) \, \sigma_s(A_2 \cap A_3)
\ee

\medskip
Proposition \ref{random-states} immediately yields
\be
\sigma_s(A_1) = (1 - \gamma)^{s-1}
\ee
From (\ref{tube6}) this gives
\be\label{tube7}
\sigma_s\left(\Phi^C(|\theta\rangle\langle\theta|) \in \tu(\rho)\right) \ge
(1 - \gamma)^{s-1} \,(1 - \sigma_s(A_2^c) - \sigma_s(A_3^c))
\ee
In order to bound $\sigma_s(A_3^c)$ we first use (\ref{prop1b}) to deduce
\be\label{tube8}
\| \Phi^C (|\psi\rangle\langle\phi|) \|_{\infty} \le
\| \Phi^C (|\psi\rangle\langle\phi|) \|_{2} \le
\| \Phi^C (|\psi\rangle\langle\theta|) \|_{2} + d\,\| | \theta \ket - | \phi \ket \|_{2}
\ee
Thus
\be\label{tube9}
\sigma_s(A_3^c) 
& = & \sigma_s\bigg\{| \theta \ket \,:\, \bigg\| \Phi^C (|\psi\rangle\langle\phi|) \bigg\|_{\infty} >
 \sqrt{\frac{6 d^2 \log d}{s}} + \sqrt{\frac{ 12 d^2\log d}{s}}\bigg\} \nonumber \\
& \le & \sigma_s\bigg\{| \theta \ket \,:\, 
\| \Phi^C (|\psi\rangle\langle\theta|) \|_{2} + d \,\| | \theta \ket - | \phi \ket \|_{2} >
 \sqrt{\frac{6 d^2 \log d}{s}} + \sqrt{\frac{ 12 d^2\log d}{s}} \bigg\} \nonumber \\
& \le & \sigma_s \bigg\{| \theta \ket \,:\, 
\| \Phi^C (|\psi\rangle\langle\theta|) \|_{2} > \sqrt{\frac{6 d^2 \log d}{s}} \bigg\} +
\sigma_s\bigg\{| \theta \ket \,:\, \| | \theta \ket - | \phi \ket \|_{2} >
\sqrt{\frac{ 12 \log d}{s}} \bigg\} \nonumber \\
& \le & (d^2 + 1) \, \Big(1 - \frac{6 \log d}{s} \Big)^{s-1} \\
\ee
where the last inequality follows from (\ref{prop2b}) and (\ref{prop1a}).

\medskip 
Turning now to $\sigma_s(A_2^c)$, note first that
\be\label{tube10}
\bigg\|  \Phi^C (|\phi\rangle\langle\phi|)- \frac{1}{d}I \bigg\|_{\infty} & \le &
\bigg\|  \Phi^C (|\phi\rangle\langle\phi|)- \Phi^C (|\theta\rangle\langle\theta|) \bigg\|_{\infty} +
\bigg\|  \Phi^C (|\theta\rangle\langle\theta|)- \frac{1}{d}I \bigg\|_{\infty} \nonumber \\
& \le &
\bigg\|  \Phi^C (|\phi\rangle\langle\phi|)- \Phi^C (|\theta\rangle\langle\theta|) \bigg\|_{2} +
\bigg\|  \Phi^C (|\theta\rangle\langle\theta|)- \frac{1}{d}I \bigg\|_{\infty} \nonumber \\
& \le & 2d \, \| | \theta \ket  - | \phi \ket \|_{2} +
\bigg\|  \Phi^C (|\theta\rangle\langle\theta|)- \frac{1}{d}I \bigg\|_{\infty}
\ee
where we used (\ref{prop1b}) for the last inequality. As in (\ref{tube9}) this gives
\be\label{tube11}
\sigma_s (A_2^c) & = & \sigma_s\bigg\{| \theta \ket \,:\, 
\bigg\|  \Phi^C (|\phi\rangle\langle\phi|)- \frac{1}{d}I \bigg\|_{\infty} >
 \sqrt{\frac{48 d^2 \log d}{s}} + 2 \, \sqrt{\frac{\log n}{n}} \bigg\} \nonumber \\
& \le & \sigma_s\bigg\{| \theta \ket \,:\, 
 \| | \theta \ket  - | \phi \ket \|_{2} > \sqrt{\frac{ 12 \log d}{s}} \bigg\} \nonumber \\
&&\hskip0.5in + \sigma_s\bigg\{| \theta \ket \,:\, 
\bigg\|  \Phi^C (|\theta\rangle\langle\theta|)- \frac{1}{d}I \bigg\|_{\infty} >
2 \, \sqrt{\frac{\log n}{n}} \bigg\} \nonumber \\
& \le & \bigg(1 - \frac{6 \log d}{s} \bigg)^{s-1} +
\sigma_s\bigg\{| \theta \ket \,:\, 
\bigg\|  \Phi^C (|\theta\rangle\langle\theta|)- \frac{1}{d}I \bigg\|_{\infty} >
2 \, \sqrt{\frac{ \log n}{n}} \bigg\}
\ee
where we used (\ref{prop1a}) for the last inequality. By assumption $\Phi \in T$, and 
therefore there is a set of input states $L$ with $\sigma_s(L) \ge 1/2$ such that
\be\label{tube12}
|\theta \ket \in L \Rightarrow 
\bigg\|  \Phi^C (|\theta\rangle\langle\theta|)- \frac{1}{d}I \bigg\|_{\infty} \le
2 \, \sqrt{\frac{\log n}{n}}
\ee
Thus
\be\label{tube13}
\sigma_s\bigg\{| \theta \ket \,:\, 
\bigg\|  \Phi^C (|\theta\rangle\langle\theta|)- \frac{1}{d}I \bigg\|_{\infty} >
2 \, \sqrt{\frac{ \log n}{n}} \bigg\} \le
\sigma_s(L^c) \le \frac{1}{2}
\ee

\medskip
Putting together the bounds (\ref{tube7}), (\ref{tube9}), (\ref{tube11}) and (\ref{tube13}) we get
\be\label{tube14}
\sigma_s(\Phi^C(|\theta\rangle\langle\theta|) \in \tu(\rho))
 &\ge & 
(1 - \gamma)^{s-1} \,\bigg(1 - \sigma_s(A_2^c) - \sigma_s(A_3^c)\bigg) \nonumber \\
& \ge &  (1 - \gamma)^{s-1} \,\bigg(1 - \bigg(1 - \frac{6\log d}{s} \bigg)^{s-1} - \frac{1}{2}
- (d^2 + 1) \, \Big(1 - \frac{6 \log d}{s} \Big)^{s-1}\bigg) \nonumber \\
& =&  (1 - \gamma)^{s-1} \,\bigg(\frac{1}{2} - (d^2 + 2) \, \Big(1 - \frac{6 \log d}{s} \Big)^{s-1}\bigg) 
\ee
The proof now follows by noting that for all $d,s \ge 2$
\be
(d^2 + 2) \, \Big(1 - \frac{6 \log d}{s} \Big)^{s-1} \le \frac{1}{4}
\ee
and hence
\be
\sigma_s(\Phi^C(|\theta\rangle\langle\theta|) \in \tu(\rho)) \ge
\frac{1}{4} \, (1 - \gamma)^{s-1} 
\ee

\section{Proof of Lemma \ref{lem:md->0}}
We use properties of the function $m_d$ derived in Section 5.7 of \cite{FKM09}.
We have
\be
m_d(y) = g(z), \quad g(z) = - \log z - (d-1) \, \log \frac{d-z}{d-1}
\ee
where $z = h^{-1}(y)$ and
\be
h(z) = z \, \log z + (d-z) \, \log \frac{d-z}{d-1} 
\ee
As was shown in \cite{FKM09} both functions $g,h$ are increasing, and $h(1)=0$.
Since the functions are analytic, their behavior near $z=1$ is determined by their power series
expansions at $z=1$. To leading order these are
\be
g(1+t) &=& \left( \frac{d}{2(d-1)} \right) \, t^2 - \left( \frac{d^2-2d}{3(d-1)^2} \right) \, t^3 + O(t^4) \nonumber \\
h(1+t)&=& \left( \frac{d}{2(d-1)} \right) \, t^2- \left( \frac{d^2-2d}{6(d-1)^2} \right) \, t^3 + O(t^4)
\ee
Setting $y=h(1+t)$, solving the second series for $t$ in terms of $y$, and substituting into the first series gives
\be
m_d(y) = y - \left( \frac{d^2-2d}{3(d-1)^2} \right) \, \left( \frac{2(d-1)}{d} \right)^{3/2} \, y^{3/2} + \cdots
\ee
Thus for sufficiently small $y$ we have $m_d(y)/y \sim 1 - k \sqrt{y}$, which implies both
statements in Lemma \ref{lem:md->0}.

\end{document}